\documentclass{article}

\usepackage{amsmath}
\usepackage{amssymb}
\usepackage{makeidx}

\usepackage{graphicx}
\usepackage{latexsym}
\usepackage{tikz}

\begin{document}

\newtheorem{theorem}{Theorem}
\newtheorem{guess1}{Proposition}
\newtheorem{guess2}{Conjecture}
\newtheorem{guess3}{Corollary}
\newtheorem{guess}{Lemma}



\makeatletter
\renewcommand{\l@section}{\@dottedtocline{1}{1.5em}{2.6em}}
\renewcommand{\l@subsection}{\@dottedtocline{2}{4.0em}{3.6em}}
\renewcommand{\l@subsubsection}{\@dottedtocline{3}{7.4em}{4.5em}}
\makeatother

\title{Removing classical singularities by use of quantum mechanical sources}

\author{Graham Weir \\
Institute of Fundamental Sciences\\
Massey University\\
Palmerston North, New Zealand \\
grahamweir@xtra.co.nz}

\label{firstpage}

\maketitle

\begin{abstract}
For distances large relative to the electron Compton wavelength, the Maxwell and gravitational fields from a bound electron in its groundstate are essentially those from a rotating, charged, massive point particle. For distances small relative to the electron Compton wavelength, the corresponding Maxwell fields and General Relativity metric, Riemann and Einstein tensors become bounded, showing that, for this example, quantum effects remove the corresponding classical singularities in electromagnetism and General Relativity. The asymptotic magnetic dipole field from the bound electron produces a constant magnetic field of several Tesla, aligned along the spin axis of the electron, at the singularity position. The corresponding apparent mass density from the gravitational field  from the bound electron is about 2kgm$^{-3}$, at the singularity position.
\end{abstract}



\section{Introduction}

Mathematical modelling can be imagined to consist of two-way abstract mappings between the world of physical measurements and the world of mathematics.  At the present time, the physical world needs to inform mathematical models of certain parameter values,\cite{Kaye&Laby} before predictions can be made. Given any physical measurement, the central hypothesis of mathematical modelling is that this measurement can be correctly and consistently predicted by a mathematical model, and additionally, successful mathematical models can correctly and consistently predict actual and hypothetical physical measurements. There are (at least) three cases when this hypothesis is challenged.

Firstly, the mathematical model is unable to correctly predict the observed measurement. This typically occurs when the mathematical model is extended beyond its limits of accuracy. A simple example is our inability to accurately predict long range weather measurements, because of our inability to perform the required numerical computations. 

Secondly, the mathematical model does not make a unique prediction. This occurs for example when a mathematical solution bifurcates, and two or more possible solutions exist. Examples of this behaviour occur when predicting thermal ignition,\cite{Balakrishnan} or alternatively, elastic or plastic failure \cite{JaegarCook69} of solid structures. Additional physical hypotheses can then select the correct branch of the solution, and the relevant prediction made, and again the central hypothesis of mathematical modelling will be upheld.

Thirdly, the mathematics can fail to make a prediction because of the occurance of a singularity in the mathematical model. Singularities appear ubiquitously in the classical theories of electro-magnetism,\cite{CorsonLorrain1962} and General Relativity,\cite{Misner-Thorne-Wheeler-73} and acceptance of these singularities has great practical utility. For example, deep theorems \cite{Hawking} \cite{Heusler} have been derived concerning the singularity associated with the emergence of our Universe from the Big Bang, or in the formation of black holes.\cite{Hawking88}

Nevertheless, there may be a serious failure in the central hypothesis of mathematical modelling, if the physical world is non-singular, while the corresponding mathematical world contains singularities. The approach of this paper is to assume that on approaching a singularity predicted by electromagnetism or General Relativity, new effects from quantum mechanics\cite{Strange2005} arise, which smear out the classical singularity and result in bounded quantities.

We demonstrate this effect with four examples: the electric potential and the magnetic dipole from a charged point particle; and the singularities in General Relativity associated with the mass and angular momentum from a point particle. We will assume quantum mechanical sources, and derive the field values in electro-magnetism at these singular points, showing that the Maxwell field is bounded there, and also show that the metric, Riemann and Einstein tensors\cite{Lawden} are bounded at the location of the corresponding classical singularity for a point particle in General Relativity.

The bound states of an electron about a nucleus is chosen as our quantum mechanical system. The well known theory of QED is available then to approximate the corresponding Maxwell fields from this bound electron. However, there is no widely accepted corresponding theory of quantum gravity, and we follow an intuitive approach to derive the corresponding metric, Riemann and Einstein tensors at the location of the classical singularity in General Relativity.

Our approach is the inverse of the Schrodinger-Poisson or Schrodinger-Newton equation,\cite{Ruiz} in which the classical gravitational potential is input, and the corresponding non-linear self-consistent Schrodinger wave function analysed. Dirac's equation for an electron has been solved in a Kerr-Newman geometry,\cite{Burinskii} but typically, singular behaviour occurs at the Compton length scale. Our approach differs from previous work, by obtaining bounded field values, at the location of classical singularities.

\section{Atomic Maxwell fields}
\label{Maxwell}

The aim of this section is to derive approximations to the Maxwell fields arising from a bound electron about a nucleus. This can be viewed as the third step in an iterative approach to solve the corresponding QED problem by: firstly approximating the Maxwell field from the nucleus using classical methods; secondly solving the corresponding Dirac equations to find approximate electron wave functions; and thirdly using these wave functions to improve approximations to the Maxwell fields about the nucleus. We call these improvements atomic Maxwell fields, because their electric current density results from quantum mechanical terms.

Our default coordinates are spherical polars. Appendix A presents the electric current density vector $J^q_a$ for the ground state of the Dirac electron. $J^q_a$ are the source terms for the vector potential $A^q_a$ in Maxwell theory,\cite{Jackson67} which obey
\begin{equation}
\label{123.003}
\Box A_a = \Box (\Phi, -{\bf A}) = \mu_0 (\rho^q c^2, -{\bf J}^q) = \mu_0 J^q_a
\end{equation}
The solution, for a (static) vector potential $A_a$, is a particular solution to (\ref{123.003}), plus the complimentary solutions. The particular solution $A_a^p$ is found by assuming that the radial and axial components are zero (because the corresponding components of $J^q_a$ are also  zero), and assuming that $A^q_a$ has the same angular dependence as the corresponding component of $J^q_a$. Since $A_t^p$ is independent of time, and $A_4^p$ is independent of the azimuthal coordinate $\phi$, the vector $A_a^p$ has zero divergence.

The approximate particular solutions $A^p_t, A^p_4$ for the vector potential satisfy
\begin{equation}
\label{123.00002}
A^p_t = R^p_t(r)  \hspace{0.5mm} _{0}Y_{0 0} ; \hspace{2mm}
A^p_4 = R^p_4(r)  \hspace{0.5mm} _{-1}Y_{1 0}
\end{equation}
\begin{equation} \nonumber
\label{123.00003}
-r^{-2}\partial_r r^2 \partial_r R^p_{t} = \frac{\mu_0 q c^2}{2 \sqrt{\pi}} (R^2 + I^2) ; \hspace{2mm}
- \partial_r r^{-2}\partial_r r^2 R^p_{4} = \sqrt{\frac{2}{3 \pi}} \mu_0 q c {\rm sgn}(m) RI
\end{equation}
where $_{s}Y_{\kappa m}$ is a spin-weighted spherical harmonic, with boundary conditions
\begin{equation} \nonumber
\label{123.00004}
\lim_{r=0}r R^p_t  = 0 = \lim_{r=\infty}R^p_t ; \hspace{2mm}
\lim_{r=0}r^2 R^p_4  = 0 = \lim_{r=\infty}R^p_4
\end{equation}

Our choice for $R^2$ and $IR$ are given in (\ref{123.ab07}), allowing us to find the approximate particular electric potential $\Phi^p$ as
\begin{equation}
\label{123.014}
\Phi^p = 
\frac{q}{4 \pi \epsilon_0} \left[
\frac{1}{r} 
 - \exp(- 2 \lambda r)
\left(\frac{1}{r} + \lambda
\right) \right]
\end{equation}
\begin{equation} \nonumber
\label{123.014z}
 = \frac{q}{4 \pi \epsilon_0} f_1(r) ; \hspace{2mm} \lim_{r \rightarrow 0} f_1 = \lambda ; \hspace{2mm} \lim_{r \rightarrow \infty} r f_1 = 1
\end{equation}
showing that  the electron charge $q$ is mostly distributed inside $2 \lambda r < 1$. From (\ref{123.014}), for small $r$, $\Phi^p \simeq$
$q \lambda [1 - (2 \lambda^2 r^2/3)]/(4 \pi \epsilon_0) $, and so the maximum of $\Phi^p$ occurs at $r = 0$. The electric field is zero at $r = 0$.

The total electric potential $\Phi$ is 
\begin{equation}
\label{123.014a}
\Phi = \Phi_Z + \Phi^p = \frac{(Z-1) e}{4 \pi \epsilon_0 r} + 
\frac{e}{4 \pi \epsilon_0}
\exp(- 2 \lambda r)
\left(\frac{1}{r} + \lambda
\right)
\end{equation}
describing a central charge at $r=0$ of $Ze$, and a screened far-field central charge of $(Z-1)e$ for large $r$, and $e$ is the magnitude of the unit electric charge ($q = -e$). The potential used in the standard model of the Hydrogen atom corresponds then to the inner limit of the total potential.

The total electric potential in (\ref{123.014a}) approximates the potential from the central nucleus and the surrounding electron. In this paper we focus only on the field from the electron, since this was derived from a quantum mechanical formulation. The singularity at $r = 0$ in (\ref{123.014a}) results from treating the nucleus classically. Presumably, if the nucleus was treated quantum mechanically, the Compton wavelength for the nucleus would smear out the singularity in the electric potential from the nucleus.

The fourth tetrad component of the particular vector potential is
\begin{equation}
\label{123.016}
A^p_4
 = \frac{\mu_0 q \hbar {\rm sgn}(m)}{8 \pi m_e} \left[\frac{1}{r^2}
- \exp(- 2 \lambda r)
\left(\frac{1}{r^2} + \frac{2 \lambda}{r} + 2 \lambda^2
\right) \right] \sin \theta
\end{equation}
\begin{equation} \nonumber
\label{123.016z}
 = \frac{\mu_0 q \hbar {\rm sgn}(m) \sin \theta}{8 \pi m_e} f_4(4)
 ; \hspace{2mm} \lim_{r \rightarrow 0} f_4 = \frac{4 \lambda^3 r}{3}
 ; \hspace{2mm} \lim_{r \rightarrow \infty} r^2 f_4 = 1
\end{equation}
which will produce a magnetic field in the radial and axial directions.

The magnetic field components $(B_r, B_2)$ can be found by writing $A^p_4$ = $S(r) \sin \theta$, and since $B = d A$,
\begin{equation}
\label{123.016a}
B = d\left[ S(r) \sin \theta r \sin \theta d \phi \right] ; \hspace{2mm}
B_r = \frac{2 S \cos \theta}{r} ; \hspace{2mm}
B_2 = - \frac{1}{r} (r S),_r \sin \theta
\end{equation}

The long distance behaviour of the magnetic field from (\ref{123.016}) - (\ref{123.016a}) is
\begin{equation}
\label{123.017}
{\bf B}
 \simeq \frac{\mu_0}{4 \pi} \frac{e \hbar {\rm sgn}(m)}{2 m_e} 
\left( \frac{2 \cos \theta}{r^3}, \frac{\sin \theta}{r^3}, 0 \right)
\hspace{4mm} \lambda r \gg 1
\end{equation}
From (\ref{123.003}),  $A^p_4 = -{\bf A}_3$. Equation (\ref{123.017}) describes a magnetic dipole, with a magnetic moment of $e \hbar m / m_e$, since $m = \pm \frac{1}{2}$.

A remarkable property of (\ref{123.016}) is that the 3D vector potential and magnetic field are always bounded, for all values or $r$, since for small $r$, $A^p_4 \simeq\mu_0 q \hbar {\rm sgn}(m) \lambda^3 r \sin \theta/(6 \pi m_e) $. The magnetic field for small $r$ is
\begin{equation}
\label{123.018}
{\bf B}
 \simeq - \frac{\mu_0 q \hbar {\rm sgn}(m) \lambda^3}{3 \pi m_e}  
\left(\cos \theta, -\sin \theta, 0 \right) ;\hspace{4mm} \lambda r \ll 1
\end{equation}
which is a constant magnetic field in the ($\pm$) vertical direction,
\begin{equation}
\label{123.018a}
B_z = \cos \theta B_r - \sin \theta B_2  = - \frac{{\rm sgn}(m) Z^3 \epsilon^4}{9} \frac{3 m_e^2 c^2}{e \hbar}; \hspace{4mm} \lambda r \ll 1
\end{equation}

The contributions to the improved vector potential come from four sources. First, the original Coulomb interaction between the nucleus and electron emerges as the zero moment solution of the homogeneous equation $\Box A = 0$. Second, higher order moment solutions to the homogeneous equation will describe the geometry of the nucleus, taking into account its finite extent.
Third, the electron charge will produce an apparent electric potential which is of bounded magnitude everywhere. Fourth, the motion of the electron will produce a distributed magnetic dipole (and higher multipole components), whose components are bounded everywhere.

The results above show that the electric potential and magnetic field from the electron are everywhere bounded.

\section{Atomic Metrics}
\label{AtomicMetrics}

Section \ref{Maxwell} showed that singularities in the classical solutions to Maxwell's Equations were eliminated when the source terms were described quantum mechanically. The aim of this section is to ask if singularites are also removed in the classical theory of General Relativity, when source terms are described quantum mechanically. We cannot expect a definitive answer to this question, since no widely accepted theory of quantum gravity exits. However, we can proceed intuitively, within the linearised theory of gravitation,\cite{Misner-Thorne-Wheeler-73} and obtain some insight into how quantum mechanics could modify the metric tensor components.
We define the corresponding metrics as atomic metrics, since their source terms result from processes at the atomic scale.

The tetrad equations we consider are
\begin{equation} 
\label{123.ab01}
T^{ab} = \rho_m v^a v^b ; \hspace{2mm} J^a = \rho_m v^a ; \hspace{2mm}
 E^{ab} = -\frac{8 \pi G}{c^4} T^{ab} ; \hspace{2mm} E^{ab}_{\hspace{3mm};b} = 0
; \hspace{2mm} \rho = \rho_m c
\end{equation}
\begin{equation} \nonumber
\label{123.ab02}
J^a = 2 \sqrt{2} m_e c \left(
\alpha^A \alpha^{\dot{A}} + \beta^A \beta^{\dot{A}}
\right)
 ; \hspace{2mm} \rho =
\frac{m_e c (R^2 + I^2)}{2} \left( 
|_{\frac{1}{2}}Y_{\kappa m}|^2 + |_{-\frac{1}{2}}Y_{\kappa m} |^2
\right)
\end{equation}
where $T^{ab}$ is our assumed energy momentum tensor corresponding to a pressureless perfect fluid, $J^a$ is the momentum flux vector for the Dirac electron given in (\ref{123.000}) - (\ref{123.ab07}), $\rho$ ($ = KE_t/c$) is the momentum density of the Dirac electron given in (\ref{123.ab01})
, $\rho_m$ mass density, and $E^{ab}$ is the Einstein tensor.

In this section we assume initially that the equations are described in a spherical polar coordinate system, with the tetrad base of $(\omega^1, \omega^2, \omega^3, \omega^4)$ = $(cdt, dr, r d\theta, r \sin \theta d \phi)$. In this tetrad, all components of the energy momentum tensor have the dimensions of energy per unit volume. For the Dirac electron, $J^2 = 0 = J^3$, and since $J^a$ and $\rho$ are functions of only $r$ and $\theta$, we have $J^b_{\hspace{1mm};b}$ = 0 = $J^b \partial_b$. The non-zero tetrad connections $\Gamma_{abc} = - \Gamma_{bac}$ follow from 
\begin{equation}
\label{123.ab03v}
\Gamma_{323} = \Gamma_{424} = \frac{1}{r} ; \hspace{2mm} \Gamma_{434} = \frac{\cot \theta}{r}
\end{equation}
and so
\begin{equation}
\label{123.ab03}
T^{ab}_{\hspace{3mm};b} = v^a J^b_{\hspace{1mm};b} + J^b \partial_b v^a + \Gamma^a_{e b} v^e J^b
= - \frac{\delta^a_3}{\rho_m} \frac{\cot \theta}{r} (J^4)^2
\end{equation}
showing that the equations in (\ref{123.ab01}) are consistent with $E^{ab}_{\hspace{3mm};b} = 0$, only if we can ignore the second order terms $(J^4)^2$. Recall that $J^1$ is a ``large component", and $J^4$ is a ``small component", making $(J^4)^2$ of second order. Consequently, we shall consider only the stress tensor components of $T^{11}$ and $T^{14}$, which are of zero and first order, respectively. The $T^{44}$ component is of second order, while the other stress tensor components in (\ref{123.ab01}) are zero.

We proceed by considering the ground state of the Dirac electron, ignoring where possible the small components,
\begin{equation}
\label{123.ab06}
J^1 \simeq \frac{m_e c R^2}{4 \pi} \simeq \rho; \hspace{2mm}
J^4 =  -J_4 \simeq - \frac{m_e c I R \sin \theta {\rm sgn}(m)}{2 \pi}
\end{equation}
\begin{equation} \nonumber
\label{123.ab08}
T^{11} \simeq \frac{m_e c^2 \lambda^3}{\pi} \exp(-2 \lambda r); \hspace{2mm}
T^{14} \simeq \frac{m_e c^2 \epsilon Z \lambda^3}{\pi} {\rm sgn}(m) \sin(\theta) \exp(-2 \lambda r)
\end{equation}

The linearised equations of General Relativity are 
\begin{equation}
\label{123.ab04r}
ds^2 = (\eta_{\mu \nu} + h_{\mu \nu})dx^{\mu} dx^{\nu} ; \hspace{2mm}
H^{ab} = h^{ab} - \frac{1}{2} h^f_{\hspace{1mm}f} \eta^{ab} ; \hspace{2mm}
\Box H^{ab} = - \frac{16 \pi G}{c^4} T^{ab}
\end{equation}
where $\eta^{ab}$ is the Minkowski metric diag$(1, -1, -1, -1)$, giving
\begin{equation}
\label{123.ab04}
- r^{-2} \partial_r r^2 \partial_r H^{11} = - \frac{16 \pi G}{c^4} T^{11}
\end{equation}
since there are no non-zero connections containing a ``1" index. From (\ref{123.ab04}) and (\ref{123.ab06}), the bounded solution for $H^{11}$, satisfies
\begin{equation}
\label{123.ab09}
H^{11} \simeq - \frac{4 G m_e}{c^2} \left[\frac{1}{r} - \left(\frac{1}{r} + \lambda \right) \exp(-2 \lambda r)\right] = - \frac{4 G m_e}{c^2} f_1
\end{equation}
where $f_1$ is defined in (\ref{123.014}).

From (\ref{123.ab06}), observe that
\begin{equation}
\label{123.ab11}
T^{14} = - \frac{\epsilon Z {\rm sgn} (m)}{2 \lambda} \partial_z T^{11}
\end{equation}
where $z$ is the vertical coordinate, and so the bounded solution for $H^{14}$ is
\begin{equation}
\label{123.ab12}
H^{14} = - \frac{2 G m_e \epsilon Z {\rm sgn} (m) \sin \theta}{\lambda c^2} f_4
\end{equation}
where $f_4$ is defined in (\ref{123.016}).

From (\ref{123.ab04r}), (\ref{123.ab09}) and (\ref{123.ab12}), the metric tensor is approximately
\begin{equation}
\label{123.ab13a}
ds^2 = \left[
1 - \frac{2 G m_e}{c^2} \left[\frac{1}{r} - \left(\frac{1}{r} + \lambda \right) \exp(-2 \lambda r)
\right]\right] c^2 dt^2
\end{equation}
\begin{equation} \nonumber
\label{123.ab13b}
- \left[
1 - \frac{2 G m_e}{c^2} \left[\frac{1}{r} - \left(\frac{1}{r} + \lambda \right) \exp(-2 \lambda r)
\right] \right] (dr^2 + r^2 d\theta^2 + r^2 \sin^2 \theta d\phi^2)
\end{equation}
\begin{equation} \nonumber
\label{123.ab13c}
+ \frac{4 G m_e \epsilon Z {\rm sgn} (m) \sin \theta}{\lambda c^2} \left[
\frac{1}{r^2} - \left(\frac{1}{r^2} + \frac{2 \lambda}{r} + 2 \lambda^2
\right) \exp(-2 \lambda r)
\right] c dt \hspace{0.5mm} r \sin \theta d\phi
\end{equation}

Comparing (\ref{123.ab13a}) with results in linearised General Relativity,\cite{Misner-Thorne-Wheeler-73} for large values of $r$, shows that
\begin{equation}
\label{123.ab13}
M = m_e ; \hspace{2mm}
S = \hbar \hspace{0.5mm}{\rm sgn} (m)
\end{equation}
and so for large $r$, (\ref{123.ab13a}) corresponds to the metric from an isolated mass $M$ equal to the electron mass, and with an angular momentum vector  of magnitude $S$ equalling $\hbar$, aligned along the $z$ axis. Thus the gravitational spin of the electron is twice its intrinsic spin. A similar factor of two arose in the early Bohr theory for the angular momentum of the electron around the nucleus. We will proceed then by using (\ref{123.ab13}). Of course, in large astronomical bodies, quantum mechanical angular momentum is insignificant relative to classical angular momentum, which is defined in terms of macroscopic length scales.

From (\ref{123.ab13a}), the metric for small $\lambda r$ is approximately
\begin{equation} \nonumber
\label{123.ab15}
ds^2 = \left[1 - \frac{2 Z \epsilon G m_e^2}{\hbar c} \right] c^2 dt^2 
- \left[1 + \frac{2 Z \epsilon G m_e^2}{\hbar c} \right] [dr^2 + r^2(d \theta^2  + \sin^2 \theta d\phi^2)]
\end{equation}
\begin{equation}
\label{123.ab16}
\hspace{5mm} + \frac{16 (Z \epsilon)^3 {\rm sgn}(m) G m_e^3 r \sin \theta}{3 \hbar^2} c dt \hspace{0.4mm} r\sin \theta d \phi
\end{equation}
showing that the speed of light is reduced, as in an ideal dielectric, with
\begin{equation}
\label{123.ab17}
n = \sqrt{
\frac{
1 + \frac{2 Z \epsilon G m_e^2}{\hbar c}
}
{1 - \frac{2 Z \epsilon G m_e^2}{\hbar c}
}
}
\end{equation}
where $n$ is the index of refraction. This effect is insignificant for electrons, but if an analogous analysis was conducted for particles of mass around the Planck mass, the speed of light would approach zero, and any particle pairs would not be able to separate. It is possible then that pair production could be significantly reduced for particles of mass around the Planck mass.

Alternatively, comparing the small $\lambda r$ behaviour in the metric from (\ref{123.ab13a}), with that from the constant density $\rho_0$ interior Schwartzschild metric, shows
\begin{equation}
\label{123.ab10}
\rho_0 = \frac{8 m_e \lambda^3}{9 \pi} \simeq 2{\rm kg m}^{\hspace{-.3mm}-3} ; \hspace{2mm}
\lambda \simeq 1.9 \times 10^{10}{\rm m}^{\hspace{-.3mm}-1} ;
\hspace{2mm}
\frac{G m_e}{c^2} \simeq 6.7 \times 10^{-58}{\rm m}
\end{equation}
The concept that black holes form at a distances of $G m_ec^{-2}$ is shown in (\ref{123.ab10}) to be invalid for masses smaller than the Planck mass, because when the black hole radius is much smaller than the corresponding Compton wave length, quantum effects alter the metric and remove the black hole horizon.

A large collection of elementary particles, with individual masses much smaller than the Planck mass, can collectively form an event horizon. This does not invalid the finiteness claim above, since event horizons do not typically occur at singular points in the Riemann tensor.

The metric in (\ref{123.ab16}) can be expressed in tetrad form, by defining
\begin{equation}
\label{123.ab18}
a = \frac{2 Z \epsilon G m_e^2}{\hbar c}
; \hspace{2mm} b =  - \frac{8 (Z \epsilon)^3 {\rm sgn}(m) G m_e^3}{3 \hbar^2}
\end{equation}
and transforming from spherical polar to cylindrical polar coordinates, using
\begin{equation}
\label{123.ab19}
l = r \sin \theta ; \hspace{2mm} z = r \cos \theta 
\end{equation}
where, in flat space, $l$ ($z$) is distance to (along) the rotation axis. Defining
\begin{equation}
\label{123.ab20}
\omega^1 = \frac{1}{\sqrt{2}} \left[
\sqrt{1-a} c dt + \left(
\sqrt{1+a} - \frac{b l}{\sqrt{1-a}}
\right)  l d\phi
\right]
; \hspace{2mm} \omega^2 = \sqrt{1+a} dl 
\end{equation}
\begin{equation} \nonumber
\label{123.ab21}
\omega^3 = \sqrt{1+a} dz ; \hspace{2mm} 
\omega^4 = \frac{1}{\sqrt{2}} \left[
\sqrt{1-a} c dt - \left(
\sqrt{1+a} + \frac{b l}{\sqrt{1-a}}
\right) l d\phi
\right]
\end{equation}
expresses the metric in (\ref{123.ab16}) as approximately
\begin{equation}
\label{123.ab22}
ds^2 = 2 \omega^1 \omega^4 - (\omega^2)^2 - (\omega^3)^2 = g_{ab} \omega^a \omega^b ; \hspace{2mm} g_{ab} = g^{ab}
\end{equation}
since we are ignoring terms in $r^2$ and higher.

The ten non-zero tetrad connections are
\begin{equation} \nonumber
\label{123.ab23}
\Gamma_{121} = - \mu = - \Gamma_{211} ; \hspace{2mm}
\Gamma_{124} = \frac{1}{2}(\mu + \nu) = - \Gamma_{214} ; \hspace{2mm}
\Gamma_{244} = \nu = - \Gamma_{424}
\end{equation}
\begin{equation}
\label{123.ab24}
\Gamma_{142} = -\frac{1}{2}(\mu - \nu) = - \Gamma_{412} ; \hspace{2mm}
\Gamma_{241} = - \frac{1}{2}(\mu + \nu) = - \Gamma_{421}
\end{equation}
\begin{equation} \nonumber
\label{123.ab25}
\beta = \sqrt{1 + a} ; \hspace{2mm} \gamma = \frac{b}{\sqrt{1-a}}
 ; \hspace{2mm}
\mu = \frac{\beta + 2 \gamma l}{2 \beta^2 l} ; \hspace{2mm}
\nu = \frac{\beta - 2 \gamma l}{2 \beta^2 l}
\end{equation}

The six non-zero two-index one forms $\Gamma_{ab}$ are
\begin{equation} \nonumber
\label{123.ab26}
\Gamma_{12} = - \mu \omega^1 + \frac{1}{2}(\mu + \nu) \omega^4 = - \Gamma_{21} 
; \hspace{2mm}
\Gamma_{14} = -\frac{1}{2}(\mu - \nu) \omega^2 = - \Gamma_{41} = -\frac{\gamma dl}{\beta}
\end{equation}
\begin{equation}
\label{123.ab27}
\Gamma_{24} =  \nu \omega^4 - \frac{1}{2}(\mu + \nu) \omega^1 = - \Gamma_{42} 
\end{equation}
while the four non-zero exterior derivatives $d \Gamma_{ab}$ follow from
\begin{equation}
\label{123.ab28}
d \Gamma_{12} =  \frac{\gamma}{2 \beta^3 l} (\omega^1 \wedge \omega^2 +
\omega^2 \wedge \omega^4) =  d \Gamma_{42} 
\end{equation}

For small $\lambda r$, the twelve non-zero Riemann tetrad components have constant magnitude, and follow from
\begin{equation}
\label{123.ab29}
R_{1212} = R_{1414} = R_{2424} = - \frac{\gamma^2}{\beta^4}
\end{equation}
For small $\lambda r$, the Einstein tensor components, are diagonal and constant
\begin{equation}
\label{123.ab30}
E_{11} = E_{22} = E_{33} = - \frac{\gamma^2}{\beta^4} = - E_{44}
= E^{22} = E^{33} = E^{44} = - E^{11}
\end{equation}
Recall that the `1' and `4' components are aligned along null vectors.\\

The constancy of the Riemann tetrad components in (\ref{123.ab29}) demonstrates that this atomic metric has removed the classical singularities implied from the behaviour of the metric for large $\lambda r$. It seems possible that such classical singularities will also be absent in a large body consisting of many bound elementary particles, because as one approaches the singularity associated with any bound elementary particle, quantum effects will smear out the classical singularities associated with that elementary particle. However, this argument does not apply to unbound elementary particles, participating in scattering processes.

The equality in magnitude of the tetrad components of the Einstein tensor in (\ref{123.ab30}) predicts stress tensor tetrad components of magnitude $|T|$
\begin{equation}
\label{123.ab31}
|T| = \left[ 
\frac{8}{9 \pi (1+a)(1-a^2)} \cdot \frac{G m_e^2}{\hbar c}
\right] 
\left[m_e c^2 \left(\frac{(Z \epsilon)^2 m_e c}{\hbar} \right)^3 \right]
\end{equation}
The right-most square bracket in (\ref{123.ab31}) equals the mass energy of the electron, within a volume based on the electron Compton wave length, divided by  $(Z \epsilon)^2$. The first non-dimensional square bracket in (\ref{123.ab31}) is incredibly small, since the non-dimensional gravitational structure constant $G m_e^2/(\hbar c) \simeq 1.8 \times 10^{-45}$.  For $Z =1$, (\ref{123.ab31}) implies that the length scale over which the electron appears to be spread, at the singularity position, is around 60km.


The expressions above in this section assumed that the electron was in its groundstate, simply because of analytic ease. However there are an infinite number of possible bound quantum states for the electron, and in principle, any one of these states could be used to derive different atomic metrics.

Atomic metrics, as derived above, are characterised by continuously differentiable metric components, varying from the inner-most constant density behaviour, to the exterior gravitational fields from isolated spinning mass sources. This differs from present classical solutions, which have specific interior and exterior solutions, joined across definite surfaces, in contrast to the continuously differentiable atomic metrics.

Finally, the inverse problem of deriving an atomic metric from an external empty space solution, is briefly discussed in Appendix B.

\section{Conclusions}

The key result of this paper was showing that the monopole and dipole electromagnetic and gravitational fields from isolated bodies yield finite field values at the singular points in classical theory, when quantum mechanical bound sources are used. Electromagnetic and gravitational field values often result from bound point sources, suggesting that both mathematically predicted and measured physical field values from bound quantum sources are indeed finite. This supports the central hypothesis of mathematical modelling that the mathematical and physical worlds should mirror properties of each other.

Our results were based on the ground state of the electron, and predicted that the bounded values at the location of the classical singularities depended on the Compton wavelength of the electron, and the fine structure constant. Since the classical field values are unbounded at the singular points, it may be thought that the bounded field values we have derived could be unrealistically high, and have little physical relevance. In contrast, the constant magnetic field value derived in (\ref{123.018a}) has magnitude
\begin{equation}
\label{1.000}
\frac{\epsilon^4 m_e^2 c^2}{3 e \hbar} \simeq 5 \hspace{1mm} {\rm tesla}
 \hspace{2mm} {\rm when} \hspace{2mm}
Z = 1 ; \hspace{2mm}
\frac{m_e^2 c^2}{3 e \hbar} \simeq 10^{9} \hspace{1mm} {\rm tesla} \hspace{2mm} {\rm when} \hspace{2mm} Z \epsilon = 1
\end{equation}
showing that for $Z$ = 1, the magnetic field value is around 5 tesla, typical of that in an ordinary NMR machine. Alternatively, in the limit of $Z \epsilon$ = 1, 
the magnetic field values in (\ref{123.018a}) and (\ref{1.000}) approach those observed in neutron stars.

The gravitational effects from the mass and spin of one electron are incredibly small. However, astronomical bodies contain an incredibly large number of fermions, which result in significant gravitational effects from large astronomical bodies. Assuming linear superposition for the gravitation field suggests that the value of the stress energy tensor in (\ref{123.ab31}) will increase linearly with the number of nucleons. The limit of this process occurs when the number of nucleons $n_c$ implies the electron is contained within the electron Compton wavelength
\begin{equation} 
\label{1.001}
n_c \simeq \frac{\hbar c}{\epsilon^6 G m_e^2} \simeq 4 \times 10^{57} ; \hspace{2mm}
M_c \simeq n_c m_n \simeq 6 \times 10^{30} \hspace{1mm} {\rm kg} \simeq
3 {\rm \hspace{1mm} solar \hspace{2mm} masses}
\end{equation}
where $m_n$ is the neutron mass. However, $M_c$ is not the Chandrasekar mass, which has the functional form $m^3_P m_n^{-2}$, where $m_P$ is the Planck mass.

The theory in this paper relating to gravitational effects was intuitive, and assumed that the stress energy tensor was dominated by the ``time-time'' component. Interestingly, from (\ref{123.ab30}) and (\ref{123.ab20}), the corresponding Einstein ``time-time'' tensor component is zero at the singularity position (being about $2 b l \gamma^2/\beta^4$).

Our intuitive analysis yielded the correct asymptotic gravitational mass, but there is uncertainty over the angular momentum analysis, which yielded a gravitational angular momentum of $\hbar$, equalling twice the intrinsic spin of $\hbar/2$ for the electron. The same angular momentum of $\hbar$ follows from a more rigourous analysis based on QED.

In contrast, we would expect an angular momentum contribution of $2 \hbar$, since the gravitational field is a spin 2 field, and angular momentum should be quantised in units of $2 \hbar$. It is well known from electro-weak theory that the electron does not occur by itself, but is coupled to the electron-neutrino, in an iso-spin doublet. If the spin of the electron-neutrino could contribute a spin of $\hbar$ to the gravitational field, then perhaps the gravitational field would be quantised in units of $2 \hbar$. This remains an open question, however, as the electron-neutrino will be in an unbound state.

Despite these uncertainties in the magnitude of some of the results, it is possible that we have captured the essence of the finite values of the corresponding metric, Riemann and Einstein tensors, at the location of the classical singularity for the linearised gravitational field.

\appendix

\section{Dirac results}

The bound states of the electron about a nucleus are described by either a pair of two spinors $(\alpha_A, \beta^{\dot{B}})$,\cite{Penrose84} or a four spinor $\psi$.\cite{Bjorken-Drell} Each predicts an electric current density $J^q_{a}$, with zero divergence. The wave functions are normalised by
\begin{equation}
\label{123.000}
\int dV J_t = m_e c^2
; \hspace{2mm}
\int dV J^q_t = q c^2 ; \hspace{2mm}
J^q_{a} = \frac{q}{m_e} J_{a}
\end{equation}
where $J_{a}$ is the mass flux vector, and $J_t$ is the corresponding time component. These equations ensure that $q$ is the total charge obtained by integrating the electric charge density over all space, for any time; and that  $m_e$ is the total mass obtained by integrating the electon mass density over all space, for any time.

In the tetrad base $(\omega^1, \omega^2, \omega^3, \omega^4)$ = $(cdt, dr, r d \theta, r \sin \theta d \phi)$, the electric current density is $J^q_a$ = $(J^q_1, 0, 0, J^q_4)$, and $J^q_1, J^q_4$ depend only on $r$ and $\theta$. Using the tetrad metric diag($1, -1, -1, -1$),
\begin{equation}
\label{123.001}
J^q_{t} = 4 q |f|^2 c^2 \left( 
|_{\frac{1}{2}}Y_{\kappa m}|^2 + |_{-\frac{1}{2}}Y_{\kappa m} |^2
\right) ; \hspace{2mm} J^q_r = 0  ; \hspace{2mm} 8 |f|^2 = R^2 + I^2
\end{equation}
\begin{equation}
\label{123.002}
J^q_{3} = 0; \hspace{2mm} J^q_{4} = - q I R c \left(
_{\frac{1}{2}}\overline{Y}_{\kappa m \hspace{1mm}-\frac{1}{2}}Y_{\kappa m} +
_{-\frac{1}{2}}\overline{Y}_{\kappa m \hspace{1mm}\frac{1}{2}}Y_{\kappa m}
\right)
\end{equation}
where $q$ is the charge ($= - e$) of the electron, $_{\frac{1}{2}}Y_{\kappa m}$ is a spin-weighted Spherical harmonic,\cite{Penrose84} and $\kappa, m$ are half-integers. $R$ and $I$ are functions of $r$ only, and are in standard form for relativistic Dirac theory.

The groundstate is $1S^{\frac{1}{2}}$, and has $\kappa = \frac{1}{2}$ and $m = \pm \frac{1}{2}$,
\begin{equation}
\label{123.001u}
|_{\frac{1}{2}}Y_{\frac{1}{2} m}|^2 + |_{-\frac{1}{2}}Y_{\frac{1}{2} m} |^2 =
\frac{1}{\sqrt{\pi}} \hspace{2mm} _{0}Y_{0 0} ; \hspace{2mm} m = \pm \frac{1}{2}
\end{equation}
\begin{equation} \nonumber
\label{123.001v}
_{\frac{1}{2}}\overline{Y}_{\frac{1}{2} m \hspace{1mm}-\frac{1}{2}}Y_{\frac{1}{2} m} +
_{-\frac{1}{2}}\overline{Y}_{\frac{1}{2} m \hspace{1mm}\frac{1}{2}}Y_{\frac{1}{2} m} = 
- {\rm sgn}(m) \sqrt{\frac{2}{3 \pi}} \hspace{1mm} _{-1}Y_{1 0}
\end{equation}
and so the source terms from the $1S^{\frac{1}{2}}$ state are
\begin{equation}
\label{123.00001}
J^q_{t} = \frac{q c^2}{2 \sqrt{\pi}} (R^2 + I^2) \hspace{0.5mm} _{0}Y_{0 0} ; \hspace{2mm}
J^q_{4} = \sqrt{\frac{2}{3 \pi}} q c {\rm sgn}(m) IR \hspace{0.5mm} _{-1}Y_{1 0}
\end{equation}

It is useful to further simplify the groundstate expressions for $R^2$ and $R I$ to
\begin{equation}
\label{123.ab07}
R^2 + I^2 \simeq
R^2 \simeq 4 \lambda^3 \exp(- 2 \lambda r) ; \hspace{2mm}
R I \simeq -2 \epsilon Z \lambda^3 \exp(- 2 \lambda r) 
; \hspace{2mm}
\lambda = \frac{m_e c Z \epsilon}{\hbar} 
\end{equation}
which are used throughout this paper, since this allows easy solution of the corresponding Maxwell fields. Here $\epsilon = e^2/(4 \pi \epsilon_0 \hbar c)$ is the fine structure constant, and $m_e$ is the electron mass.

\section{The Kerr atomic metric}

The Kerr metric\cite{Kerr} in Boyer-Linquist coordinates\cite{BoyerLinquist} is
\begin{equation}
\label{8.497d}  \nonumber
ds^2 = \left(1 - \frac{m r^2}{\rho^2}\right)c^2 dt^2 - \frac{\rho^2}{\Delta} dr^2 - \frac{\rho^2}{r^2} r^2 d \theta^2
- \left(1 + \frac{a^2}{r^2} + \frac{m a^2 \sin^2 \theta}{\rho^2}\right) r^2 \sin^2 \theta d \phi^2
\end{equation}
\begin{equation}
\label{8.497c}
+ \frac{2 s r^2 \sin \theta}{\rho^2} r \sin \theta d\phi c dt; \hspace{2mm}
\rho^2 = r^2 + a^2 \cos^2 \theta; \hspace{1mm} \Delta = r^2 + a^2  - \frac{2GM r}{c^2} ;
\end{equation}
\begin{equation} \nonumber
\label{8.497daa}
a = \frac{S}{M c} ; \hspace{2mm} m = \frac{2 G M}{c^2 r}
; \hspace{2mm} s = \frac{2 G S}{c^3 r^2}
; \hspace{2mm} \left(\frac{a}{r} \right)^2 = \left( \frac{s}{m}
\right)^2
\end{equation}
corresponding to the steady external gravitational field from an isolated mass $M$, with angular momentum $S$ along the positive $z$ coordinate.

Given an external classical metric such as the Kerr metric in Boyer-Lindquist coordinates in (\ref{8.497c}), there is not a unique way in which this can be extended to an atomic metric, although there is a natural extension. For an atomic metric, from (\ref{123.ab09}), the mass terms are multiplied asymptotically by $r^{-1}$, whereas from (\ref{123.ab12}), the angular momentum terms are multiplied asymptotically by $r^{-2}$. From (\ref{8.497c}), the components $g_{ef}$ of the Kerr metric tensor (in tetrad form) can be written as $g_{ef}$ = $g_{ef}(m, s, a, r, \theta)$, and then the functions in (\ref{123.ab09}) and (\ref{123.ab12}) can be used to produce the atomic metric naturally associated with the Kerr metric in Boyer-Lindquist coordinates, through the substitutions
\begin{equation} \nonumber
\label{8.497daar}
m \rightarrow \frac{2 G M f_1}{c^2}
; \hspace{2mm} s \rightarrow \frac{2 G S f_4}{c^3}
; \hspace{2mm} \left(\frac{a}{r} \right)^2 \rightarrow \left( \frac{a f_4}{f_1}
\right)^2
\end{equation}

The resulting atomic metric reduces to the Kerr metric asymptotically, and for small $\lambda r$, the metric is bounded. However, the expression $a f_4$ becomes linear in $r$ for small $\lambda r$. Consequently, the terms containing $a^2$ in the metric are quadratic in $r$ for small $\lambda r$, and so are insignificant. To within a constant (near unity) scaling of the $r$ coordinate, the atomic Kerr metric tensor then reduces to that from the linear theory, given in (\ref{123.ab16}), for small $\lambda r$.\\

\printindex
   \label{lastpage}
\end{document}